\documentclass[aps,prx,superscriptaddress,twocolumn,showpacs,longbibliography]{revtex4-1}

\usepackage{amssymb}
\usepackage{amsmath}
\usepackage{amsthm}
\usepackage{graphicx}
\usepackage{amsfonts}
\usepackage{color}
\usepackage{times}
\usepackage{natbib}
\usepackage{comment}
\usepackage{soul}
\usepackage{booktabs}
\usepackage{pifont}
\usepackage{hyperref}
\hypersetup{
        colorlinks=true,
}

\newcommand{\bra}[1]{\langle #1|}
\newcommand{\ket}[1]{|#1\rangle}
\newcommand{\average}[1]{\langle #1\rangle}

\newcommand{\ketbra}[2]{| #1 \rangle \langle #2 |}

\DeclareMathOperator{\tr}{tr}

\newcommand{\beq}{\begin{equation}}
\newcommand{\eeq}{\end{equation}}
\newcommand{\bea}{\begin{eqnarray}}
\newcommand{\eea}{\end{eqnarray}}
\newcommand{\ba}{\begin{align}}
\newcommand{\ea}{\end{align}}

\newcommand{\hW}{\widehat{W}}

\begin{document}

\title{No-go theorem for the characterisation of work fluctuations in coherent quantum systems}

\author{Mart{\'i} Perarnau-Llobet}
\email{marti.perarnau@icfo.es} 
\affiliation{ICFO-Institut de Ciencies Fotoniques, The Barcelona Institute of Science and Technology, 08860 Castelldefels, Barcelona, Spain}

\author{Elisa B\"aumer}
\email{ebaeumer@itp.phys.ethz.ch}
\affiliation{ICFO-Institut de Ciencies Fotoniques, The Barcelona Institute of Science and Technology, 08860 Castelldefels, Barcelona, Spain}
\affiliation{Institute for Theoretical Physics, ETH Zurich, 8093 Z\"urich, Switzerland}

\author{Karen V. Hovhannisyan}
\email{karen.hovhannisyan@icfo.es} 
\affiliation{ICFO-Institut de Ciencies Fotoniques, The Barcelona Institute of Science and Technology, 08860 Castelldefels, Barcelona, Spain}

\author{Marcus Huber}
\email{marcus.huber@univie.ac.at} 
\affiliation{Departament de F\'isica, Universitat Aut\`onoma de Barcelona, 08193 Bellaterra, Spain} 
\affiliation{Institute for Quantum Optics and Quantum Information (IQOQI), Austrian Academy of Sciences, Boltzmanngasse 3, A-1090 Vienna, Austria}

\author{Antonio Ac\'in}
\email{antonio.acin@icfo.es} 
\affiliation{ICFO-Institut de Ciencies Fotoniques, The Barcelona Institute of Science and Technology, 08860 Castelldefels, Barcelona, Spain} 
\affiliation{ICREA, Pg. Llu\' is Companys 23, 08010 Barcelona, Spain}

\begin{abstract}

An open question of fundamental importance in thermodynamics is how to describe the fluctuations of work for quantum coherent processes. In the standard approach, based on a projective energy measurement both at the beginning and at the end of the process, the first measurement destroys any initial coherence in the energy basis. Here we seek for extensions of this approach which can possibly account for initially coherent states. We consider all measurement schemes to estimate work and require that (i) the difference of average energy corresponds to average work for closed quantum systems, and that (ii) the work statistics agree with the standard two-measurement scheme for states with no coherence in the energy basis. We first show that such a scheme cannot exist. Next, we consider the possibility of performing collective measurements on several copies of the state and prove that it is still impossible to satisfy simultaneously requirements (i) and (ii). Nevertheless, improvements do appear, and in particular we develop a measurement scheme which acts simultaneously on two copies of the state and allows to describe a whole class of coherent transformations.

\end{abstract}

\maketitle

\section{Introduction}

The second law of thermodynamics, as a statement about average work and average heat, remains correct even when one goes down to the microscopic scale. Nevertheless, unlike the macroscopic case, fluctuations of work and heat become significant for small systems, and are not negligible anymore. As a consequence, and starting with the seminal papers \cite{b_k,jarz}, fluctuations of work have become a topic of central interest to statistical thermodynamics (see, e.g., \cite{massimo,michele,hanggirep,udo}).

At the same time, small scales bring quantum effects along with them, and the very notion of work variable becomes challenging to define \cite{talkner2007,ll5,review,Armen2005,Aberg2013,Horodecki2013,david,Gentaro,GentaroII,alla,solinas,Anders,Saar,Gallego2015,tanggi,jon,Nelly,Hayashi,AndersII}. Indeed, it is of no surprise that although quantum mechanics is very definitive when it comes to averages (hence average work is a well-defined quantity), it abolishes the notion of phase-space trajectories, thereby making it impossible to define the work variable by directly applying the classical intuition. This problem is generic to quantum mechanics, and is captured by the so-called full counting statistics \cite{nazarov,Patrick}. In fact, the latter can be used in the problem of defining a work variable \cite{utsumi,solinas,solinasII}.

In this article, the scenario under consideration consists of a system described by a quantum state $\rho$ and Hamiltonian $H$. The system undergoes an externally controlled Hamiltonian evolution, described by a unitary transformation $U$, and ends up in a new quantum state, $\rho\xrightarrow{\textrm{evol}} U\rho U^\dagger$, with a new Hamiltonian $H'$. Given this process, there are several approaches to obtaining the statistics of work, namely, the set of outcomes $\{ W \}$ and their probability distribution $P_W$ \cite{review,alla,tanggi}. This diversity comes from the fact that, unlike in classical mechanics, in order to build $P_W$ in quantum physics one has to specify the measurement scheme through which such knowledge is obtained. Furthermore, measurements are invasive, so that the observation itself can modify the original process, $\rho\xrightarrow{\textrm{evol}} U\rho U^\dagger$, and hence the energetics. 

In order to design a scheme that is minimally invasive, and at the same time physically well-motivated, we demand two requirements on the corresponding $P_W$.

\emph{(i) In a closed quantum system, the difference of average internal energy corresponds to work.} This imposition goes back to the very definition of work and heat in phenomenological thermodynamics in which, for closed systems, every change of energy comes in form of work. For the considered process, this is equivalent to demanding
\begin{equation}
\sum_{W} W P_W= \tr\left( U\rho U^{\dagger} H' \right) - \tr(\rho H).
\label{AvW}
\end{equation}
This should remain valid for all $\rho$'s and $U$'s.

\emph{(ii) For states with no quantum coherence, the results of classical stochastic thermodynamics should be recovered.} Classical stochastic thermodynamics, in particular fluctuation theorems, have been extended in the quantum regime by the two-projective-energy-measurements scheme \cite{talkner2007,jorge,hal,michele}, referred to as TPM scheme here. Here we demand strict agreement with this scheme for classical diagonal states. By this requirement we ensure that our definition of fluctuating work has a proper classical limit \cite{Saar,tanggi}.

While these two requirements appear reasonable, it is straightforward to see that the existing definitions of work do not satisfy both of them. For example, the TPM scheme trivially satisfies (ii), but fails to satisfy (i) whenever the state has quantum coherence, as the first measurement becomes invasive and destroys all the coherences in the state \cite{fstlaw}. The incompatibility remains also for Gaussian energy measurements \cite{tanggi,FN:UntouchedWork}. On the other hand, the operator of work~\cite{Armen2005} satisfies (i) but not (ii). Other recent definitions of work ~\cite{alla,solinas,AndersII}, in which both requirements are satisfied, suffer from negative probabilities, which cannot be understood as a quantum measurement \cite{muck}.

The main result of this paper is to prove rigorously that this incompatibility is not just a shortcoming of particular approaches, but rather a fundamental limitation imposed by quantum mechanics. Namely, we show that there exists no measurement of work that satisfies simultaneously the two requirements imposed above for all processes and states. This shows that observing the micro-statistics inherently changes the global (average) work when dealing with quantum systems. This result represents a no-go result on the definition of work as a fluctuating quantity in quantum mechanics, and sheds light into different definitions of work in the literature  \cite{Gentaro,GentaroII,alla,talkner2007,solinas,tanggi,Hayashi,AndersII}.

Besides this no-go result, we also construct new schemes for estimating fluctuating work which can approximately describe coherent transformations. More concretely, we construct a scheme that satisfies (ii) exactly, and (i) to a certain level of approximation. The main idea behind the scheme is to use global measurements, where a number of copies of the state independently undergoing the same process can be measured simultaneously. As such, the back action of the measurement can be reduced, and hence we can work more closely with the original process $\rho\xrightarrow{\textrm{evol}} U\rho U^\dagger$.  This represents a first step towards the measurement of fluctuating work in quantum coherent evolutions.

\section{Fluctuations of work, generalized quantum measurements, and convexity}

In this work we assume that fluctuations of work can be characterised by a real random variable W, to which a probability distribution $P_W$ can be assigned \cite{FootnoteQuasiprobabilities}. We also follow the standard approach, adopted in most of the previous attempts, and assume that work fluctuations can be observed. In quantum physics, this means that they can be estimated through a measurement process, which in turn can always be described by a generalized quantum measurement, defined by a Positive-Operator-Valued Measure (POVM) \cite{CommentMeas}. A POVM is a set of non-negative Hermitian operators $\{M^{(W)}\}$, which satisfy $\sum_{\{ W \}} M^{(W)} = \mathbb{I}$. Each possible value of work $W$ is associated with an operator $M^{(W)}$, so that the probability to obtain $W$ can be computed through the generalized Born rule:
\begin{equation}
P_W=\tr\left(\rho M^{(W)}\right).
\label{P(W)}
\end{equation}
We consider measurement operators $M^{(W)}$ that can depend on the process, $\Pi=(H,H',U)$, but are independent of the initial state $\rho$: 
\begin{equation}
M^{(W)} = M^{(W)}(\Pi).
\label{M(W)}
\end{equation}
Indeed, one would like to have a \textit{universal} scheme to estimate work so that there is no need for adjusting the measurement apparatus to the initial state.

One may question why quantum work fluctuations should correspond to an observable quantity and, thus, be defined though a measurement. Interestingly, it is possible to arrive at expressions \eqref{P(W)} and \eqref{M(W)} using an alternative, slightly more formal approach. The starting point is the same, namely work fluctuations should be described by a random variable, where to each outcome $W$, a probability $P_W$ is assigned. In general, this assignment can depend both on the process and the state: $P_W=P_W(\Pi,\rho)$. Now, it is natural to assume that if one picks as initial state $\rho_1$ with probability $p_1$ and state $\rho_2$ with probability $p_2$ ($p_1+p_2=1$), then the resulting work distribution \emph{is the mixture of the individual work distributions}, $\{ P_W(\Pi,\rho_1) \}$ with probability $p_1$ and $\{ P_W(\Pi,\rho_2) \}$ with probability $p_2$. In other words,
\begin{equation} \label{linearity}
P_W(\Pi,p_1\rho_1+p_2\rho_2) = p_1P_W(\Pi,\rho_1) + p_2P_W(\Pi,\rho_2)
\end{equation}
for all $W$s. Imposing this requirement, a Gleason-type argument (see Appendix~\ref{App:Convexity}) guarantees that for each $W$ there exists a non-negative Hermitian operator $M^{(W)}$ independent of $\rho$, such that $P_W(\Pi,\rho)=\tr(M^{(W)}\rho)$. Thereby, this shows that invoking POVMs and imposing \eqref{P(W)} and \eqref{M(W)} can interchangeably be replaced with the single linearity condition \eqref{linearity}. Put differently, \eqref{P(W)} and \eqref{M(W)} not only imply linearity of $P_W$ with respect to convex combinations of density matrices, but are also equivalent to it.

\section{Minimal requirements for the statistics of work}
\label{Sec:MinimalRequirements}

Given the previous definitions, we can now express the requirements presented in the introduction in detail. Regarding requirement (i), the average work of a certain process is given by $\sum_W \tr\left(M^{(W)} \rho\right) W$. By introducing the operator
\begin{equation}
X=\sum_W W M^{(W)},
\label{defX}
\end{equation}
it can be rewritten as $\average{W}_{\rho}=\tr{(X \rho)}$ \cite{footav}. From expression \eqref{AvW}, one then obtains, $\tr{(X \rho)}= \tr\left((H-U^{\dagger} H'U)\rho \right)$. Since this must hold for any $\rho$, requirement (i) is equivalent to
\begin{equation}
X=H-U^{\dagger} H'U.
\label{condX}
\end{equation}
Note that this does not fix the measurement scheme -- there can be many combinations of non-negative $M^{(W)}$s summing up to $\mathbb{I}$ and yielding the same $X$.

In order to describe requirement (ii), let us briefly recall the TPM scheme. Expand the Hamiltonians as $H=\sum_i E_i \ketbra{i}{i}$, and $H'=\sum_i E_i' \ketbra{i'}{i'}$ \cite{FNFiniteDim}. Now, the first step of the scheme consists of a projective energy measurement of $\rho$, which yields $E_i$ with probability $\bra{i}\rho \ket{i}$. Only after this measurement, the process is implemented, and the state $\ket{i}$ evolves under $U$. Finally, a projective energy measurement with respect to the final Hamiltonian is performed, yielding $\ket{j'}$ with conditional probability $|\bra{j'}U \ket{i}|^2$. To this realization, a work value $W^{(ij)}=E_i-E'_j$ is assigned, with the corresponding probability of occurrence $p^{(ij)}=\rho_{ii} \hspace{1mm} p_{i,j}$, where $p_{i,j}= |\bra{j'}U \ket{i}|^2$. The resulting probability distribution for work can be written as $P_\textrm{TPM}(W)= \sum_{ij} \delta(W-W^{(ij)}) p^{(ij)}$, where $\delta$ is the Dirac delta function. As noted in \cite{augusto}, the whole scheme can be expressed by the following POVM: $M^{(W)}_\textrm{TPM}=\sum_{ij} \delta(W-(E_i-E'_j)) p_{i,j} \ketbra{i}{i}$. Formally, requirement (ii) then simply states that
\begin{equation}
\tr(\rho M^{(W)})=\tr(\rho M^{(W)}_\textrm{TPM}), \hspace{7mm} \forall W, \hspace{3mm}
\forall \rho=D_H(\rho)
\label{condwiii}
\end{equation}
where $D_H$ is the operation removing all coherence between eigenspaces of $H$.

Before proving our main result, the incompatibility of these two requirements, let us study condition \eqref{condwiii} in more detail. Generally speaking, realizations of work ($W$ in  $M^{(W)}$) can take any real value. However, by considering $\rho=\ketbra{k}{k}$ $\forall k$ in \eqref{condwiii}, and setting $W\neq E_i-E'_j$, we obtain
\begin{align}
\langle k | M^{(W)} | k\rangle =0 \quad  \forall k \quad {\rm if} \quad  W\neq E_i-E'_j.
\label{pij0}
\end{align}
Since $M^{(W)}$ is a non-negative operator, this means that $M^{(W)}=0$, whenever $W\neq E_i-E'_j$. Hence, the only values of $W$ that can be observed, i.e., those for which $M^{(W)}\neq 0$, are the energy differences. 

Next, we focus on the case where the possible values of work $E_i-E'_j$ are non-degenerate. We introduce the operators $M^{(ij)} \equiv M^{(E_i-E'_j)}$, and write the POVM of the TPM scheme as 
\begin{align}
M^{(ij)}_{\rm TPM}= p_{i,j} \ketbra{i}{i}.
\label{MijTPMdd}
\end{align}
Consequently, \eqref{condwiii} will acquire the following form:
\begin{align}
\tr(\rho M^{(ij)})=\rho_{ii} \hspace{1mm} p_{i,j} \quad \quad \forall \rho=D_H(\rho)\;\; \text{and}\;\; \forall i,j.
\label{pij}
\end{align}
By again considering $\rho=\ketbra{k}{k}$ $\forall k$, we obtain from \eqref{pij} that $\bra{k}M^{(ij)}\ket{k}=\delta_{ik} p_{i,j}$. Now, since there is only one non-zero diagonal element, the non-negativity of $M^{(ij)}$ implies that all off-diagonal elements are zero. Therefore, the conditions \eqref{pij0} and \eqref{pij} unambiguously fix the measurement operators $M^{(ij)}$ to be identical to the ones in \eqref{MijTPMdd}.

\section{No-go result for the characterization of work fluctuations in coherent processes}

We are now ready to prove that the two requirements cannot be jointly satisfied for all processes and states. For that, note that it is enough to construct a counter-example. Consider a two-level system with initial state $\rho$. It starts with Hamiltonian $H=\epsilon \ketbra{1}{1}$ and ends up with $H'=\epsilon' \ketbra{1}{1}$, and the process is such that the unitary evolution operator is given by $U=\ketbra{0}{+}+\ketbra{1}{-}$, with $\ket{\pm}=(\ket{0}\pm\ket{1})/\sqrt{2}$. As we showed above, requirement (ii) fixes the POVM matrices to be
$M^{(ij)}= p_{i,j} \ketbra{i}{i}$, which, through (\ref{defX}), give us an expression for $X$: $X=-\epsilon' \frac{\ketbra{0}{0}}{2}+(2\epsilon-\epsilon') \frac{\ketbra{1}{1}}{2}$. 
On the other hand, requirement (i) demands through \eqref{condX} that $X$ equals $H-U^{\dagger} H'U =\epsilon \ketbra{1}{1}-\epsilon' \ketbra{-}{-}$. For any nonzero $\epsilon'$, the two expressions for $X$ do not coincide. Hence, this provides the counterexample.

This no-go result shows that any apparatus for measuring work, that gives correct classical outputs for classical states, necessarily disturbs the process so much that it changes the average work. The implications of this result for existing methods to describe the fluctuations of work in externally driven quantum systems are discussed in Table ~\ref{tableA}.

\section{Extension to global measurements}

In order to reduce the back-action of the measurements, we now extend our considerations to global measurements, where $N$ copies of the state independently undergoing the same process can be globally processed. In this case, expression \eqref{P(W)} is replaced by,
\begin{equation}
P_W=\tr\left(\rho^{\otimes N} M^{(W)}\right).
\label{P(w)n}
\end{equation}
Examples of global measurements include sequential measurements, in which a different measurement is implemented in each copy, 
\begin{equation}
M^{(W)}=M^{(W)}_1\otimes M^{(W)}_2 \otimes ... \otimes M^{(W)}_N,
\label{Mwseq}
\end{equation}
feedback-measurements, in which $M^{(W)}_{j}$ can depend on the previous outcomes, and finally entangling measurements, which cannot be written as a convex combination of measurements like \eqref{Mwseq}. Clearly, global measurements can provide an advantage here, and the intuition behind this is two-fold: On the one hand, one can measure some copies at the beginning and some others at the end of the process, thereby minimizing the disturbance induced by the measurement apparatus. On the other hand, in the many-copy case the relative weight of energy-basis coherences becomes less significant~\cite{matteo}. It is also important to note that by assuming the form \eqref{P(w)n}, we break the convexity \eqref{linearity} of $P_W$, thereby increasing the class of allowed functions. 

When considering $N$ copies of the state, $\rho^{\otimes N}$, there are two natural ways to generalize our previous considerations: Either one considers the total work extracted in the process $\rho^{\otimes N} \longrightarrow (U\rho U^{\dagger})^{\otimes N}$, or one coarse-grains the  measurements to estimate the work extracted from a single copy. In the latter case, the other $N-1$ copies are used to obtain a more refined description of the evolution. In either case, we show that no measurement scheme exists that can simultaneously satisfy (i) and (ii) exactly, and thereby extend our previous result to collective measurements. For clarity of the discussion, here we focus on the individual work, and leave the details of the total work for Appendix~\ref{App:TotalWork}.

For global measurements on $N$ copies of the state, the operators $M^{(W)}_N$ act on $\rho^{\otimes N}$ instead of $\rho$. Then, requirement (ii) can be expressed as $\tr(\rho^{\otimes N} M^{(W)})=\tr(\rho M^{(W)}_{\rm TPM})$ $\forall \rho=D_H(\rho)$. Requirement (i) reads as $\tr(\rho^{\otimes N} X)=\tr \left(\rho H\right) - \tr\left( U \rho U^{\dagger} H'\right)$, $\forall \rho$, where $X=\sum_{W} W M^{(W)}$. Notice that essentially the same restrictions are imposed on the measurement operators $M^{(W)}$, which now act on a Hilbert space of dimension $d^N$ instead of $d$, the dimension of $\rho$. This gives an enormous freedom that was not present before.

Nevertheless, despite the freedom to choose the $M^{(W)}$, we construct a process where both requirements cannot be simultaneously satisfied (see Appendix~\ref{App:IndividualWork}). The counterexample is based on taking unitaries of the form $U(\varepsilon)=\sqrt{1-\varepsilon^2} \mathbb{I} + \varepsilon i \sigma_y$, to then show that, if $\varepsilon$ decreases fast enough with the increase of $N$, the fluctuations arising from $U(\varepsilon)$ can never be completely characterized. Hence we show the incompatibility between preserving the average work and recovering the classical limit for the most general conceivable measurements.

\begin{table}[]
\centering
\begin{tabular}{@{}lccc@{}}
\toprule[1pt]
                            & Measurable\hspace{4mm} & Fluct. theor. \hspace{4mm} & Coherent proc. \\ \midrule[0.5pt]
\multicolumn{1}{l}{TPM scheme}       &                \ding{51}      &   \ding{51}                   &         \ding{55}           \\ \cmidrule(r){1-1}
\multicolumn{1}{l}{Operator of work} &        \ding{51}                  &             \ding{55}         &            \ding{51}        \\ \cmidrule(r){1-1}
\multicolumn{1}{l}{Quasiprobabilities}                  &              \ding{55}            &                \ding{51}      &              \ding{51}      \\ \bottomrule[1pt]
\end{tabular}
\caption{Comparison between three different approaches to
characterize the fluctuations of work in externally driven quantum systems: the TPM scheme \cite{talkner2007}, the operator of work \cite{Armen2005}, and approaches based on quasiprobabilities \cite{alla,solinas,AndersII}. Each approach fails to satisfy a different requirement, as expected from the no-go result.}
\label{tableA}
\end{table}

\section{A new measurement scheme to evaluate the quantum fluctuations of work}

Based on the idea of collective measurements, here we construct a new measurement scheme to approximately describe the fluctuations of work in coherent processes. For that, let us first introduce
\begin{align}
T_j\equiv U^{\dagger}|j' \rangle \langle j'| U,
\end{align}
where we recall that $H'= \sum_j E_j' |j' \rangle \langle j'| $. Consider now the expansion, $T_j = T_j^{\rm (diag)}+T_j^{\rm (off-diag)}$, with $T_j^{\rm (diag)}=\sum_k |\langle j'| U | k \rangle|^2 \hspace{1mm}|k\rangle \langle k|$ and $T_j^{\rm (off-diag)} = \sum_{l\neq s} \langle l| U | j' \rangle \langle j'| U | s \rangle \hspace{1mm}|l\rangle \langle s|$. Clearly, $T_j^{\rm (off-diag)}$ acts on the off-diagonal elements of $\rho$, and, since  $\tr(U \rho U^{\dagger} H' ) =\sum_j E_j' \tr(\rho T_j)$, it brings the coherent part of work.  

Now, the measurement scheme acts on two copies of $\rho$, $\rho^{\otimes 2}$, and is given by the following POVM elements (see Appendix~\ref{App:MS} for a detailed derivation),
\begin{align}
M^{(ij)}_{\lambda}=\ketbra{i}{i} \otimes \left(\bra{i} T_j^{\rm (diag)} \ket{i} \mathbb{I} + \lambda
T_j^{\rm ( off-diag)} \right),
\label{GeneralMij}
\end{align}
where the parameter $\lambda$ is chosen such that
\begin{equation}
\lambda=\max_{\alpha} (\hspace*{0.5mm}\alpha \hspace*{1mm}: \hspace*{1mm}M^{(ij)}_{\alpha}\geq 0 \hspace*{2mm}\forall i,j).
\label{lambdaen}
\end{equation}
The probability $\tr(\rho^{\otimes 2} M^{(ij)}_{\lambda})$ is then associated with the value of work $E_i-E_j'$. 

The measurement scheme \eqref{GeneralMij} is a combination of two measurements: A projective energy measurement on the first copy of $\rho$ at the beginning of the process, and a (in general) non-projective measurement on the second copy after being evolved through $U$. The parameter $\lambda$ given by \eqref{lambdaen} is introduced to ensure the positivity of the POVM elements, so that this measurement scheme is operationally well defined and can be experimentally implemented. 
Furthermore,  notice that
\begin{align}
M^{(ij)}_{\lambda}=M^{(ij)}_{\rm TPM} \otimes \mathbb{I} + \lambda \ketbra{i}{i} \otimes T_j^{\rm off-diag}.
\label{GeneralMijII}
\end{align}
Hence the scheme can be seen as an extension of the standard TPM scheme: It acts in the same way on the diagonal part of $\rho$, and additionally brings information about the coherent work through the second term in \eqref{GeneralMijII}. More precisely, the enhancement with respect to the TPM scheme is quantified by $\lambda$: For $\lambda=1$ the average work remains unchanged, whereas for $\lambda=0$ one obtains the same results of the TPM scheme. In Appendix~\ref{App:Qubtis}, we determine $\lambda$ for generic qubit evolutions.

In order to show the power of this scheme, we focus on a particular family of evolutions, namely, maximally coherent processes, which are unitary operations of the form
\begin{align}
W=\frac{1}{\sqrt{d}}\sum_{j,k}^{d-1} e^{-\frac{2 \pi i}{d}jk} \ket{j} \bra{k},
\label{UexAs}
\end{align}
where $d$ is the Hilbert space dimension. Unitary operations of the form \eqref{UexAs} map basis states to maximally coherent states and vice versa, and hence are of great importance here. For such processes, the maximization \eqref{lambdaen} yields $\lambda=1$, see Appendix~\ref{App:MaximallyCoherent}. Furthermore, the POVM elements take the simple form 
\begin{equation}
M^{(ij)}_{\lambda=1}=\ketbra{i}{i} \otimes W^{\dagger} \ket{j} \bra{j} W,
\label{Mijex}
\end{equation}
which simply corresponds to a projective energy measurement on the first copy, followed by a projective energy measurement on the second copy after the evolution. 

Let us now look at the probabilities generated by \eqref{Mijex} for the simplest instance of the evolution \eqref{UexAs} with $d=2$ acting on a fully coherent state, i.e.,
\begin{align}
&|+\rangle \overset{W}{\longrightarrow} \ket{0}.
\label{processExample}
\end{align}
with $|+\rangle= (\ket{0}+\ket{1})/\sqrt{2} $. By applying  \eqref{Mijex} on $|+\rangle^{\otimes 2}$, and using $W|0\rangle=W^{\dagger}|0\rangle= \ket{+}$, one obtains $p^{(00)}=p^{(10)}=1/2$ and $p^{(01)}=p^{(11)}=0$. This predicts that the probability of ending in the ground state, $p^{(10)}+p^{(00)}$, is 1. These results are in contrast with those predicted with the TPM scheme, given by $p^{(00)}=p^{(01)}=p^{(10)}=p^{(11)}=1/4$, which bear little resemblance to the factual evolution.

\section{Conclusions}

Our results show that two physically necessary properties of quantum work, namely, respecting the classical limit and obeying the first law of thermodynamics, cannot be simultaneously measured. As a consequence, while the observation of  work fluctuations does not change the work output for macroscopic processes, this is no longer true in quantum systems with coherence. This result sheds light on the crucial role of measurements \cite{Hayashi,jacobs,yikim,cyril,philipp,Alonso,kais} and coherence \cite{lostaglio,cwiklinski,raam,mitchinson,Misra,Fusco} in quantum thermodynamics, and seems to imply that there will probably never be an equivalently universal notion of a work variable that is independent of the context in quantum mechanics.

The basic reason behind this incompatibility is the presence of quantum coherence, together with the back action induced by quantum measurements. In order to decrease the back action, we explored the possibility of using collective measurements. Although we showed that the no-go result remains valid for such global measurements, the set of describable coherent transformations increases. In particular, using a measurement on two copies of the state, we provided a new scheme that can approximately describe the fluctuations in quantum coherent processes. 

Future work also includes a comparison between the methods developed here for describing the fluctuations of work in coherent processes and other approaches in the literature \cite{Gentaro,GentaroII,alla, solinas,AndersII,Hayashi,tanggi,solinasII,nazarov,Patrick,utsumi,johan,alvaro}. Particularly interesting are also the results on fluctuations of work obtained in the context of the resource theory of thermodynamics, where the fluctuations of work are directly mapped upon the state of an external work-exchange agent -- the "weight" \cite{johan,alvaro}. As a final remark, we note that the scheme \eqref{GeneralMij} can be used to approximately characterize the fluctuations of work in work extraction processes from entangled states \cite{us}.

\section*{Acknowldegments}

We thank Markus M\"uller for suggesting the use of collective measurements during the COST conference in Porquerolles. We further thank Armen Allavherdyan, Peter H\"anggi, and Peter Talkner for their useful comments on the manuscript. This work is supported by the Swiss NSF (AMBIZIONE PZ00P2\_161351), the ERC CoG QITBOX, an AXA Chair in Quantum Information Science, the Spanish MINECO (Project No.~FIS2013-40627-P and FOQUS FIS2013-46768-P, Severo Ochoa grant SEV-2015-0522 and Grant No.~FPU13/05988), Fundacion Cellex and the Generalitat de Catalunya (SGR875). MH would like to acknowledge funding from the Austrian Science Fund (FWF) through the START project Y879-N27. All authors are grateful for support from the EU COST Action MP1209 ``Thermodynamics in the quantum regime''.

\newpage

\appendix

\section{Linearity and POVMs} \label{App:Convexity}

Here we prove the main claim of the passage around Eq.~\eqref{linearity}. As is explained there, we are willing to impose the condition
\bea \label{poki1}
P_W(\Pi,\sum_i p_i\rho_i)=\sum_ip_iP_W(\Pi,\rho_i),
\eea
where $p_i\geq0$, $\sum_ip_i=1$, and $\rho_i$ are quantum states, to hold for the work distribution. This is a natural and hence a highly desirable property or any probability distribution associated to a physical (operational) property of a system (or a process). The idea is that $\sum_i p_i\rho_i$ describes a statistical mixture of ensembles described by $\rho_i$s, and taken, respectively, with probabilities $p_i$. This picture suggests that if while estimating $P_W(\Pi,\sum_i p_i\rho_i)$ one comes across the ensemble described by $\rho_i$, the output will be $\{ P_W(\Pi,\rho_1) \}$, and this happens with probability $p_i$. Hence, the overall output will be the RHS of \eqref{poki1}.

Now, for $d=\dim \mathcal{H}_S\geq 3$, where $\mathcal{H}_S$ is the Hilbert space of the system, it is easy to see that no extra assumption is necessary to prove that \eqref{poki1} entails Eqs.~\eqref{P(W)} and \eqref{M(W)}. Indeed, take $\sum_i\mathcal{E}_i=\mathbb{I}$ to be an arbitrary rank-1 projective resolution of identity in $\mathcal{H}_S$ (i.e., an orthonormal basis). In that case, all $\mathcal{E}_i$s are states, so we have
\bea
\sum_i P_W(\Pi,\mathcal{E}_i) = d \sum_i \frac{1}{d} P_W(\Pi,\mathcal{E}_i) =
\\
=d P_W\left(\Pi,\sum_i \frac{1}{d}\mathcal{E}_i \right)=d P_W\left(\Pi,\frac{\mathbb{I}}{d}\right).
\eea
Following the terminology of Ref.~\cite{gleason}, this means that $P_W$ is a frame function with weight $d P_W\left(\Pi,\frac{\mathbb{I}}{d}\right)\geq 0$. Therefore, the Gleason's theorem \cite{gleason} applies here directly, and implies that, for any pure state $|\psi\rangle$,
\bea \label{poki2}
P_W(\Pi,|\psi\rangle\langle\psi|)=\langle \psi|M^{(W)}(\Pi)|\psi\rangle,
\eea
where $M^{(W)}$ is a non-negative Hermitian operator. By using \eqref{poki1}, it is straightforward to extend \eqref{poki2} to
\bea \label{poki3}
P_W(\Pi,\rho)=\tr(M^{(W)}(\Pi)\rho),
\eea
for any state $\rho$. Finally, the observation that $\sum_W P_W=1$ necessitates $\tr\left(\rho \sum_W M^{(W)} \right)=1$ to hold for any $\rho$, and hence implies that $\sum_W M^{(W)}=\mathbb{I}$, completes the proof of the statement in the main text.

The case of $\dim \mathcal{H}_S =2$ is a little more subtle, and, for the proof below to hold, we need to both extend the domain of $P_W$ to all $0\leq \rho \leq \mathbb{I}$ and require \eqref{poki1} on the whole extended domain of $P_W$. Additionally, we would need $P_W$ to be a continuous function of $\rho$, which, given the physical setting at hand, is a natural assumption, as small changes in initial state should entail in small changes in output work. 

To proceed, let us take an arbitrary $N$-element POVM $\{M_\alpha\}$ and calculate
\bea
\sum_\alpha P_W(M_\alpha)=N\sum_\alpha\frac{1}{N}P_W(M_\alpha)=
\\
=NP_W\left(\frac{\sum_\alpha M_\alpha}{N}\right) = NP_W\left( \frac{\mathbb{I}}{N} \right).
\eea
Formulated otherwise, as long as $N$ is fixed, $P_W$ is a frame function for $N$-component POVMs, with a non-negative weight \cite{caves}. Now, to complete the proof, it suffices to notice that this makes $P_W$ a frame function for, say, trine measurements, and, consequently, the corresponding result in \cite{caves} ensures that $P_W(\Pi,\rho)$ is of the form \eqref{poki3}.

\section{Collective measurements}

\subsection{Total work}
\label{App:TotalWork}

In this section, we consider the total work extracted from $N$ copies of a state, $\rho^{\otimes N}$, each of them undergoing a unitary evolution $U$, i.e., 
\begin{equation}  
\rho^{\otimes N} \longrightarrow (U\rho U^\dagger)^{\otimes N}.
\end{equation}
 As described in the main text, we would like to find a POVM, $\{M^{(W)}\}$, so that the corresponding work distribution $P_W$
\begin{description}
\item[(i)] yields as an average work the change of average energy of the $N$ copies,  for all $ \rho$'s and $U$'s, and
\item[(ii)] agrees with the TPM probabilities if $\rho$ is diagonal.
\end{description}
Importantly, here we consider the \emph{total} work extracted from the $N$ copies. In what follows, we describe each condition in detail, and show how they become incompatible for most evolutions $U$ and states $\rho$. 
For simplicity of the arguments, we focus on the case of qubit evolutions. Note that this is not a restriction, as all we need is a counterexample. 

We first introduce some notation. Let ${\bf k} $ be  the $n$-bit string ${\bf k} = k_1 \cdots k_N$, with $|{\bf k} | = \sum_{i} k_i$ being the Hamming weight (number of 1s) of the string. The states $\ket{{\bf k} } = \ket{k_1}\cdots\ket{k_N}$ run over all $2^N$ energy eigenstates of the total Hamiltonian, 
\begin{align*}
H^{\rm (total)}=\sum_{j=1}^n H_j,
\end{align*}
 with $H_j=\mathbb{I}^{\otimes (j-1)} \otimes H \otimes \mathbb{I}^{\otimes (N-j)}$. The state $\ket{\bf k}$ has energy $E_k=\sum_iE_{k_i}$, and similarly, we define final states $\ket{\bf l'}$ with energy $E_l'=\sum_iE_{l_i}'$.  
We will interchangeably use the vector notation or the explicit indices, i.e.,  
\begin{equation*}
\bra{{\bf i}} \gamma \ket{{\bf j} } = \gamma_{\bf i,j } = \gamma_{i_1...i_N,j_1...j_N}
\end{equation*}
and notice  also,
\bea
\big(\rho^{\otimes N}\big)_{\bf i,j} \equiv \rho^{\otimes N}_{\bf i,j}=\prod_n\rho_{i_nj_n}.
\label{propappe}
\eea 
Through the text, we will also use  permutations $\{ \sigma_k \}$, which act on the vectors ${\bf i}$,  so that $\sigma_k ({\bf i})$ is a permutation of ${\bf i}$. Note that $\sum_{\sigma_k}=\frac{N!}{(N-|{\bf i}|)! |{\bf i}|!}$.

Let us now study condition (i). It can be written as
\beq
\langle W\rangle=\sum_W W \tr(M^{(W)}\rho^{\otimes N})=N \tr((H-U^\dagger H' U)\rho),
\nonumber
\eeq
where $H$ and $H'$ are the initial and final Hamiltonians, respectively, and $U$ is the unitary evolution operator.
Let us again define $X=\sum_W W M^{(W)}$ and $\hW =H-U^\dagger H' U$. Thus we can rephrase the condition as:
\beq
\tr\left(\rho^{\otimes N} X\right)=N\tr(\rho\hW)=\tr\left(\rho^{\otimes N}\sum_m\hW_m\right) \qquad \forall \rho,
\nonumber
\eeq
where $\hW_m=\mathbb{I}^{\otimes (m-1)}\otimes \hW\otimes\mathbb{I}^{\otimes(N-m)}$. Element-wise, this condition reads,
\begin{align}\label{elementwise}
&\sum_{\bf i,j}\left(\prod_n\rho_{i_nj_n}\right)X_{\bf j,i} \nonumber\\&
=\sum_{\bf i,j}\left(\prod_n\rho_{i_nj_n}\right)\left(\sum_m \hW_{j_mi_m}\prod_{k\neq m}\delta_{j_ki_k}\right),
\end{align}
where we used \eqref{propappe}. Since (\ref{elementwise}) must hold for all $\rho$, we can take the derivative of both sides over $\rho_{\bf ab}$, $\partial \rho_{\bf ab}=\prod_i\partial\rho_{a_ib_i}$, obtaining:
\begin{equation}
\sum_{\sigma_i}X_{\sigma_i(\bf a),\sigma_i(\bf b)}=\sum_{\sigma_i}\sum_m\hW_{b_ma_m}\prod_{k\neq m}\delta_{b_ka_k},
\nonumber
\end{equation}
where we are taking the sum over all permutations $\sigma_i$. Picking some $a_1\neq b_1$, $a_i=b_i $ $\forall i\geq 2$ yields:
\bea\label{xw}
\sum_{\sigma_i}X_{\sigma_i(a_1,...,a_N),\sigma_i(b_1,a_2,...,a_N)}=\frac{N!}{(N-|{\bf a}|)! |{\bf a}|!}\hW_{b_1a_1}.
\nonumber
\eea

Let us now study the TPM-induced constraints (ii).  
Now, for any diagonal $\rho$, $\rho_{diag}$, condition (ii) reads as 
\begin{align*}
P_W&=\tr(\rho_{diag}^{\otimes N}M^{(W)})=\tr(\rho_{diag}^{\otimes N}M^{(W)}_{\rm TPM})
\end{align*}
which implies, by recalling the definition  $M^{(kl)} \equiv M^{(E_l'-E_k)}$, that,
\begin{align*}
\tr(\rho_{diag}^{\otimes N}M^{(kl)})=\sum_{\sigma_i} \rho_{\sigma_i({\bf k}),\sigma_i({\bf k})} \sum_{\sigma_j} p_{\sigma_i({\bf k}),\sigma_j({\bf l})} 
\end{align*}
where $p_{\bf k, l}$ is the transition probability of going from the state $\ket{\bf k}$ to the state $\ket{\bf l}$. Since we are dealing with
processes of the form $\rho^{\otimes N} \longrightarrow U^{\otimes N} \rho^{\otimes N} U^{\dagger \otimes N}$, 
we have that $p_{\sigma_i({\bf k}),\sigma_j({\bf l})} = p_{\sigma_a({\bf k}),\sigma_b({\bf l})} $, $\forall \sigma_i,\sigma_j,\sigma_a,\sigma_b$. 
Hence we have that
\begin{align}\label{tpmconstraintsII}
\tr(\rho_{diag}^{\otimes N}M^{(kl)})&=\rho_{\bf k, k} p_{\bf k,l}  \sum_{\sigma_i} \sum_{\sigma_j} 
=\rho_{\bf k, k} p_{\bf k,l} C_{kl},
\end{align}
where 
\begin{align*}
C_{kl}=\frac{N!}{(N-|{\bf k}|)! |{\bf k}|!} \frac{N!}{(N-|{\bf l}|)! |{\bf l}|!},
\end{align*}
and  $p_{\bf k,l} =|\bra{{\bf l}} U^{\otimes N} \ket{{\bf k}}|^2$, $\rho_{\bf k,k}= \bra{\bf{k}}\rho^{\otimes N} \ket{\bf{k}}$,  $\ket{\bf{k}}$ is any state with energy $E_k$, and similarly $\ket{\bf{l}}$ is any state with energy $E_l'$. 
Writing both the left and right hand side of \eqref{tpmconstraintsII} explicitly, we obtain,
\begin{align}
\sum_{\bf t} M^{(kl)}_{\bf t,t}\prod_n\rho_{t_nt_n} \nonumber\overset{!}{=}C_{kl} \left(\prod_n \rho_{k_n,k_n}\right) p_{\bf k,l}.
\end{align}
Again, this must hold for all $\rho_{ diag}$'s. Hence we can take the derivative of both sides over $\rho_{\bf a,a}$, where $\partial \rho_{\bf a,a}=\prod_i\partial\rho_{a_ia_i}$, obtaining:
\begin{align}
&\sum_{\sigma_i} M^{(kl)}_{\sigma_i(\bf a),\sigma_i(\bf a)}
=C_{kl} \sum_{\sigma_i}\left(\prod_n\delta_{\sigma_i(k_n),a_n}\right)p_{\bf k,l}.~~~~~~~
\nonumber
\end{align}
This indicates that non-zero diagonal elements can only appear at matrix elements with indices of the form $M_{\sigma_i({\bf k}),\sigma_i({\bf k})}$ $\forall \sigma_i$. In order to relate it to \eqref{xw}, note that this implies  that non-diagonal non-zero elements can only occur at $M_{\sigma_i(k_1...k_N),\sigma_j(k_1...k_N)}$ $\forall \sigma_i, \sigma_j , i\neq j$. However, $M_{\sigma_i(a_1...a_N),\sigma_i(b_1a_2...a_N)}$ is not of that form if $a_1\neq b_1$, which means that the left hand side in Eq.~\eqref{xw} is zero. This is a contradiction, as we can always find some $a$, $b$ and $U$ such that $\hW_{b_1a_1}\neq 0$. This concludes the proof.

\subsection{Individual work}
\label{App:IndividualWork}

Now we move to the case of individual work, in which $N-1$ copies are used to gain a better description of the coherent evolution of a single copy.
In this case, requirement (i) reads,
\begin{align}
&\tr\left(\rho^{\otimes N} X\right)=\tr \left(\rho H\right) - \tr\left( U \rho U^{\dagger} H'\right)  \hspace{10mm} \forall \rho
\label{condiglobalII}
\end{align}
where $X=\sum_{W} W M^{(W)}$; whereas requirement (ii) takes the form,
\begin{align}
\tr\left(\rho_{\rm diag}^{\otimes N} M^{(ij)}\right)=\tr\left(\rho_{\rm diag} M^{(ij)}_{\rm TPM}\right) \hspace{5mm} \forall \rho_{\rm diag}.
\label{condiglobal}
\end{align}

By focusing again on qubit systems, let us consider unitary operations of the form,
\begin{equation}
U(\epsilon)=\sqrt{1-\epsilon^2} \mathbb{I} + \epsilon i \sigma_y
\label{Uepsilonn}
\end{equation}
with $\epsilon>0$, and cyclic processes, where $H'=H=\ketbra{1}{1}$. For the unitary \eqref{Uepsilonn} and the state $\rho_{\rm diag}=p_0 \ketbra{0}{0}+p_1\ketbra{1}{1}$, condition \eqref{condiglobal} can be expressed as,
\begin{equation*}
\sum_{{\bf k}} p_1^{|{\bf k} |} p_0^{N-|{\bf k} |} \bra{{\bf k}}M^{(ij)} \ket{{\bf k}}= \epsilon^2 ( \delta_{i0} p_0+ \delta_{i1} p_1).
\end{equation*}
From this expression, it is clear that,  $\bra{{\bf k} }M^{(ij)} \ket{{\bf k}} \leq \epsilon^2$  $\forall {\bf k}$. Because the operators $M^{(ij)}$ must be positive, this condition implies that the (free) off-diagonal terms of $M^{(ij)}$ must satisfy,
\begin{equation}
\bra{{\bf k}} M^{(ij)} \ket{{\bf l} } \leq \epsilon^2 \hspace{10mm} \forall {\bf k },{\bf l }.
\label{condoffd}
\end{equation}
Consider now the average work through the measurement scheme,
\begin{equation*}
X=\sum_{ij} W^{(ij)} M^{(ij)}=M^{(10)}-M^{(01)}
\end{equation*}
and the state,
\begin{equation}
\ket{+}=\frac{1}{\sqrt{2}}(\ket{0}+\ket{1}).
\label{+exjj}
\end{equation}
Using \eqref{condoffd} we can then obtain the following bound,
\begin{align}
\left| \tr\left((\ketbra{+}{+})^{\otimes n}X\right) \right|=\left| \frac{1}{2^{n}}\sum_{{\bf k,l }}  \bra{{\bf k }}X \ket{{\bf l }} \right| < 2^{N+1} \epsilon^2.
\label{boundappp}
\end{align}

Let us know consider requirement (i). One finds, for the evolution \eqref{Uepsilonn}, that, 
\begin{align*}
\average{W}_{\rho}&=\tr\left( \rho(H-U^{\dagger}(\epsilon)HU(\epsilon)) \right)\nonumber\\ &=\epsilon^2 (\rho_{11}-\rho_{00})-2\epsilon \sqrt{1-\epsilon^2} {\rm Re}[\rho_{01}]
\end{align*}
In particular, for the state \eqref{+exjj},
\begin{equation*}
\left|\average{W}_{\ket{+}}\right| =  \epsilon \sqrt{1-\epsilon^2}
\end{equation*}
Now, if we choose $\epsilon=1/(N2^{N+1})$, we obtain from \eqref{boundappp}, for $N$ large enough that, 
\begin{align*}
\left| \tr\left((\ketbra{+}{+})^{\otimes N}X\right) \right|< \frac{1}{N^2 2^{N+1}} < \frac{1}{N 2^{N+1}}  \approx \left|\average{W}_{\ket{+}}\right|
\end{align*}
This finishes the proof: If $\epsilon$ in \eqref{Uepsilonn} is small enough, then there exists no measurement scheme compatible with \eqref{condiglobal} that can satisfy \eqref{condiglobalII}. This generalizes the result of the main text to arbitrary collective measurements, i.e., measurements that act on a finite number of copies of $\rho$.

\section{A measurement scheme to describe the quantum fluctuations of work}
\label{App:MS}

In this section we construct a measurement scheme to characterise the fluctuations of work in coherent processes for individual work.  We consider generic evolutions, $\rho \rightarrow U \rho U^{\dagger}$ and $H \rightarrow H'$, where we recall the definitions, 
$H=\sum_i E_i \ketbra{i}{i}$, 
$H'=\sum_i E_i' \ketbra{i'}{i'} = \sum_i E_i' V\ketbra{i}{i}V^{\dagger}$, where $V$ transforms the Hamiltonian basis, $\ket{i'}=V\ket{i}$. It will be convenient to introduce the unitary operator,
\begin{equation*}
U'=V^{\dagger} U.
\label{OpW}
\end{equation*}
Let us also recall that a POVM is a set of Hermitian operators
$\{M^{(W)}\}$, which satisfy 
\begin{align}
M^{(W)} \geq 0 \quad \text{and} \quad \sum_W M^{(W)} = 1.
\label{condijad}
\end{align}

Consider the POVM elements $M^{(ij)}$. For reasons that will become clear through the proof, we take the following ansatz,
\begin{equation}
M^{(ij)}= \ketbra{i}{i} \otimes S^{(ij)},
\label{MijAg}
\end{equation}
where we note that the $S^{(ij)}$'s are functions of the elements of $U$ and are still to be fixed.
The operators $M^{(ij)}$ satisfying conditions  \eqref{condijad} implies that
\begin{align}
\sum_j S^{(ij)} =\mathbb{I},
\label{Tidentity}
\\
 S^{(ij)} \geq 0.
\label{Tpositivity}
\end{align}

When dealing with two copies of $\rho$ requirement (ii) reads as,
\begin{align}
&\tr(\rho_{\rm diag}^{\otimes 2} M^{(ij)})=\bra{i} \rho_{\rm diag} \ket{i}  \hspace{1mm} |U'_{ji}|^2,  \hspace{7mm} \forall \rho_{\rm diag}
\label{pijexII}
\end{align}
Inserting \eqref{MijAg} into \eqref{pijexII}, we obtain $\tr\left(\rho_{\rm diag}S^{(ij)}\right)= |U'_{ji}|^2$, $\forall \rho_{\rm diag}$. This suggests the following ansatz for the operators $S^{(ij)}$,
\begin{equation}
S^{(ij)}= |U'_{ji}|^2 \mathbb{I} + T_j^{\rm (off-diag)}
\label{Tijoff}
\nonumber
\end{equation}
where $ T_j^{\rm (off-diag)}$ is a matrix made up of off diagonal elements only. We have freedom to choose $ T_j^{\rm (off-diag)}$ up to the constraints \eqref{Tidentity} and \eqref{Tpositivity}.

Now, by recalling that $X= \sum_{ij} (E_i -E_j') M^{(ij)}$, 
we can compute the average work  obtained through this measurement scheme as, $\tr(\rho^{\otimes 2} X) $,
\begin{align*}
\tr(\rho^{\otimes 2} X_{i}) &= \sum_{ij} E_i \tr\left(\rho^{\otimes 2} M^{(ij)}\right)- \sum_{ij} E_i' \tr\left(\rho^{\otimes 2} M^{(ji)}\right)
\end{align*}
Let us compute each term individually. For the first one we obtain,
\begin{align*}
\sum_{ij} E_i \tr\left(\rho^{\otimes 2} M^{(ij)}\right)&=\sum_{i} E_i  \tr\left(\rho^{\otimes 2} \ketbra{i}{i} \otimes \mathbb{I} \right) \nonumber\\
&=\sum_i E_i \bra{i} \rho \ket{i} =\tr(\rho H)
\end{align*}
which gives us the initial average energy. For the second term,
\begin{align}
 \sum_{ij} E_i' &\tr\left(\rho^{\otimes 2} M^{(ji)}\right)=\nonumber\\
 &= \sum_i E_i'  \tr\left(\rho^{\otimes 2} \left( \sum_j \ketbra{j}{j} \otimes S^{(ji)}\right)\right)
 \nonumber\\
 &= \sum_i E_i' \sum_j \bra{j} \rho \ket{j}  \tr\left(\rho S^{(ji)}\right)
 \nonumber\\
 &= \sum_i E_i' \sum_j \bra{j} \rho \ket{j}  \tr\left( \left( |U'_{ij}|^2+ \tr(\rho  T_i^{\rm (off-diag)}) \right)\right).
 \label{Eisecond}
\end{align}
On the other hand, we can compute the average energy change (i.e., Eq.~\eqref{AvW}),
\begin{align}
&\tr\left(\rho (H - U^{\dagger} H' U)\right) \nonumber\\
&= \sum_i E_i \bra{i} \rho \ket{i} -  \sum_i E_i' \sum_{jk} U'_{ij} U_{ik}^{'*} \bra{j} \rho \ket{k} \nonumber\\
&=  \sum_i E_i \bra{i} \rho \ket{i} -\sum_i E_i'  \bigg(\sum_{j} \bra{j} \rho \ket{j} |U'_{ij}|^2 \nonumber\\
&\hspace{5mm}+ \sum_{l\neq k} U'_{il} U_{ik}^{'*} \bra{l} \rho \ket{k} \bigg)
\label{Xiigeneral}
\end{align}
where in the last equality we reordered some terms and changed some indexes for convenience. 

From inspecting equations \eqref{Eisecond} and \eqref{Xiigeneral}, we infer that the choice,
\begin{align}
T_i^{\rm (off-diag)}=\sum_{l\neq k} U'_{il} U_{ik}^{'*}\ket{k} \bra{l}
\label{Aijfc}
\end{align}
leads to $\tr(\rho (H-U^{\dagger}H' U))=\tr(\rho^{\otimes 2} X)$, as desired. Explicitly, we obtain that the operators $S^{(ij)}$ take the form
\begin{equation}
S^{(ij)}= |u_{ji}|^2 \mathbb{I} + \sum_{l\neq k} U'_{jl} U_{jk}^{'*}\ket{k} \bra{l}
\label{Tijfc}
\nonumber
\end{equation}
However, at the moment this is just a formal choice: In order to obtain a proper quantum measurement conditions \eqref{Tidentity} and \eqref{Tpositivity} need to be satisfied --and, in fact, due to our previous no-go result, we know that this is not possible for all evolutions. Regarding \eqref{Tidentity}, we obtain,
\begin{align*}
\sum_j S^{(ij)}&=\sum_j |U'_{ji}|^2 \mathbb{I} +\sum_{l\neq k}  \left( \sum_j  U'_{jl} U_{jk}^{'*} \right) \ket{k} \bra{l} \nonumber\\
&=\mathbb{I} +\sum_{l\neq k}  \delta_{kl}  \ket{k} \bra{l}=\mathbb{I},
\end{align*}
where we used that $U^{'\dagger} U' =\mathbb{I}$. Hence, our choice naturally satisfies constraint \eqref{Tidentity}. The positivity constraint \eqref{Tpositivity} will depend on the particular choice for $U$. In order to ensure positivity, we introduce the parameter $\lambda$, and define, 
\begin{align}
S^{(ij)}_{\lambda} = |U'_{ji}|^2 \mathbb{I} +\lambda T_j^{\rm (off-diag)}
\label{TijfcII}
\end{align}
where $\lambda \in [0,1]$ is chosen according to $\lambda=\max_{\alpha}  (\alpha \hspace{2mm} | \hspace{2mm} S^{(ij)}_{\alpha}\geq 0 \hspace{1mm} \forall i,j )$. This, together with \eqref{Aijfc}, leads to Eq.~\eqref{GeneralMij}.

\section{Maximally coherent processes}
\label{App:MaximallyCoherent}

Let us now apply the general considerations of the last section to processes that can generate maximal coherence, or conversely extract work from maximally coherent states. In particular, we consider unitary operations of the form,
\begin{align}
U=\frac{1}{\sqrt{d}}\sum_{j,k}^{d-1} e^{-\frac{2 \pi i}{d}jk} \ket{j} \bra{k}
\label{UexA}
\end{align}
where $d$ is the dimension of the Hilbert space. We also consider cyclic processes. In this case, setting $\lambda=1$, we can easily obtain the operators $S^{(ij)}$ in \eqref{Tijfc}, and they take the form
\begin{align}
S^{(ij)}&=\frac{1}{d}\mathbb{I}+\frac{1}{d} \sum_{l\neq k}^{d-1} e^{-\frac{2 \pi i}{d} j(l-k)} \ket{k}\bra{l}
\nonumber\\
&=\frac{1}{d} \sum_{l,k=0}^{d-1} e^{-\frac{2 \pi i}{d} j(l-k)} \ket{k}\bra{l}=U^{\dagger} \ket{j} \bra{j} U,
\end{align}
Clearly, in this case the operators $S^{(ij)}$ are positive, and hence the choice $\lambda=1$ in the measurements \eqref{TijfcII} is well justified. Recall that this implies that the second requirement (ii) can be satisfied exactly in this process.

\section{Application: Fluctuations of generic qubit coherent evolutions}
\label{App:Qubtis}

Let us now exemplify the potential of collective measurements by focusing on measurements performed on two copies of a qubit, undergoing a coherent evolution. Here, the POVM elements from Eq.~\eqref{GeneralMij}, take the form
\[ M^{(00)}= \left( \begin{array}{cccc}
|U'_{00}|^2& \lambda U_{00}^{'*} U'_{01} & 0 & 0\\
 \lambda U_{01}^{'*} U'_{00}&|U'_{00}|^2 & 0 & 0 \\
0 & 0 & 0 & 0 \\
0 & 0 & 0 & 0
\end{array} \right),\]
\[ M^{(01)}= \left( \begin{array}{cccc}
|U'_{10}|^2 & \lambda U_{11}^{'*} U'_{10} & 0 & 0\\
\lambda U_{10}^{'*} U'_{11}&|U'_{10}|^2 & 0 & 0 \\
0 & 0 & 0 & 0 \\
0 & 0 & 0 & 0
\end{array} \right),\]
\[ M^{(10)}= \left( \begin{array}{cccc}
0& 0  & 0 & 0\\
0 &0 & 0 & 0 \\
0 & 0 & |U'_{01}|^2 & \lambda U_{00}^{'*} U'_{01}\\
0 & 0 & \lambda U_{01}^{'*} U'_{00}& |U'_{01}|^2
\end{array} \right),\]
\[ M^{(11)}= \left( \begin{array}{cccc}
0& 0  & 0 & 0\\
0 &0 & 0 & 0 \\
0 & 0 & |U'_{11}|^{2} & \lambda U_{11}^{'*} U'_{10}\\
0 & 0 & \lambda  U_{10}^{'*} U'_{11}& |U'_{11}|^2
\end{array} \right).\]
Let us now use that  $U'$ can be parametrised, up to phase shifts, as
\[ U'= \left( \begin{array}{cc}
\cos\alpha& -\sin\alpha  \\
 \sin\alpha &\cos\alpha
\end{array} \right).\]
Then, these operators become positive if one takes
\begin{equation*}
\lambda= \frac{\min\{\cos^2 \alpha, \sin^2 \alpha\}}{|\cos{\alpha} \sin{\alpha}|}.
\label{xyex}
\end{equation*}

Assume now, for simplicity, that $\alpha \in [0, \pi/4]$, so that $\lambda= \tan \alpha$. Then, by explicitly computing the transition probabilities $p^{(ij)}={\rm Tr} (\rho^{\otimes 2}M^{(ij)})$, we obtain:
\begin{align*}
p^{(00)}&=\rho_{00} (\cos^2(\alpha) - 2 \sin^2 (\alpha) \hspace{1mm}{\rm Re}(\rho_{01})) 
\nonumber\\
p^{(01)}&=\rho_{00} (\sin^2(\alpha) + 2 \sin^2(\alpha) \hspace{1mm}{\rm Re}(\rho_{01})) 
\nonumber\\
p^{(10)}&=\rho_{11} (\sin^2(\alpha) - 2 \sin^2(\alpha) \hspace{1mm} {\rm Re}(\rho_{01})) 
\nonumber\\
p^{(11)}&=\rho_{11} (\cos^2(\alpha) + 2 \sin^2(\alpha) \hspace{1mm} {\rm Re}(\rho_{01})).
\end{align*}
Note that, whereas the first term brings information about the diagonal elements (in fact, it gives the probabilities as obtained by the TPM scheme), the second term in $p^{(ij)}$s brings information about the off-diagonal elements of $\rho$.


\begin{thebibliography}{99}
\bibitem{fstlaw} On the other hand, the TPM scheme satisfies \eqref{AvW} with $\rho$ substituted by its decohered version. In other words, the TPM scheme satisfies the energy balance, but for the process it generates. However, the latter is generally different from the original one.
\bibitem{FN:UntouchedWork} In \cite{tanggi}, for Gaussian energy measurements in the von Neumann measurement model, it is observed that, when the pointer is weakly coupled to the system, the average work is given by \eqref{AvW}; and, for strong coupling, one recovers the TPM results. This motivates the authors to name the quantity \eqref{AvW}  untouched work,  as it refers to the average work in absence of any back action from the measurement apparatus.
\bibitem{footav} Here $\langle\cdot\rangle$ denotes the average, and the subscript $\rho$ is to indicates that the systems starts in state $\rho$.
\end{thebibliography}
\end{document}